\documentclass[a4paper,12pt]{iopart}

\usepackage{iopams}
\usepackage{graphicx}
\usepackage[colorlinks=true,linkcolor=blue,citecolor=blue,urlcolor=blue]{hyperref}
\usepackage{bbm}
\usepackage{multirow}

\usepackage{tocloft}

\newcommand{\ub}{{\bi u}}

\newcommand{\be}{\begin{equation}}
\newcommand{\ee}{\end{equation}}
\newcommand{\bea}{\begin{eqnarray}}
\newcommand{\eea}{\end{eqnarray}}

\usepackage{color}

\renewcommand{\mailto}[1]{{\href{mailto:#1}{\tt #1}}}

\begin{document}

\title[Relativistic BEC]{Relativistic Bose--Einstein Condensates: a new\\system for analogue models of gravity}

\author{S Fagnocchi, S Finazzi, S Liberati, M Kormos and A Trombettoni}
\address{SISSA via Beirut 2-4, I-34151 Trieste, Italy and INFN, Sezione di Trieste, Italy}
\eads{\mailto{serena.fagnocchi@sissa.it}, \mailto{finazzi@sissa.it}, \mailto{liberati@sissa.it}, \mailto{kormos@sissa.it} and \mailto{andreatr@sissa.it}}

\begin{abstract}
In this paper we propose to apply the analogy between gravity and condensed matter physics to relativistic Bose--Einstein condensates (RBECs), i.e. condensates composed by relativistic constituents.
While such systems are not yet a subject of experimental realization, they do provide us with a very rich analogue model of gravity,
characterized by several novel features with respect to their non-relativistic counterpart. Relativistic condensates exhibit
two (rather than one) quasi-particle excitations, a massless and a massive one, the latter disappearing in the non-relativistic limit. We show that the metric associated with the massless mode is a generalization of the usual acoustic geometry allowing also for non-conformally flat spatial sections. This is relevant, as it implies that these systems can allow the simulation of a wider variety of geometries.
Finally, while in non-RBECs the transition is from Lorentzian to Galilean relativity, these systems represent an emergent gravity toy model where Lorentz symmetry is present (albeit with different limit speeds) at both low and high energies. Hence they could be used as a test field for better understanding the phenomenological implications of such a milder form of Lorentz violation at intermediate energies.

\end{abstract}

\submitto{\NJP}
\noindent Accepted: 25 Mar 2010.
\newpage

\setcounter{tocdepth}{2}
\section*{Contents}
\vspace{-30pt}
\tableofcontents

\section{Introduction}
Analogies in physics have often provided deep insight and inspiration to deal with fundamental problems. In the last 20 years, {\it analogue models of gravity}~\cite{LRR} had this role with respect to pressing issues in gravitation theory, such as the mechanism of Hawking radiation, the fate of Lorentz invariance at ultra-short distances and the nature of spacetime and gravity as possibly emergent phenomena.

The general idea behind analogue models of gravity is that in many physical systems it is possible to identify suitable excitations that propagate as fields on a curved spacetime. In particular, there is a wide class of condensed matter systems that admit a hydrodynamical description within which it is possible to show that acoustic disturbances propagate on an effective geometry, the so called `acoustic metric'~\cite{unruh81}. In fact, the propagation of excitations in the hydrodynamical regime can be described through a relativistic equation of motion in a curved spacetime.  So even starting from non-relativistic equations one can show that excitations are endowed with a Lorentz invariant dynamics where the Lorentz group is generally associated with an invariant speed coinciding with the speed of sound. Furthermore, these systems generically show a breakdown of such acoustic regime leading to Lorentz violations at high energies (as expected from the Newtonian nature of the fundamental equations). The richness and power of this analogy have led in recent years to several applications.

There is nowadays a vibrant activity toward the realization in laboratories of analogue black holes~\cite{Tachnion} and toward the observation of the associated Hawking radiation~\cite{hawking}. Acoustic metrics showing black hole horizons are found in many of the different systems proposed to realize this analogy: phonons in weakly interacting Bose--Einstein condensates~\cite{BEC}, Fermi gases~\cite{fermi}, superfluid Helium~\cite{He}, slow light~\cite{slow}, non-linear electromagnetic waveguides~\cite{waveguide} and ion rings~\cite{ions}. In~\cite{P-G} a possible mapping to transform the acoustic metric into a metric conformally equivalent to the Schwarzschild one has been pointed out. There are attempts to create also some analogue of rotating black hole geometries, with the main aim to predict super-radiance of the modes~\cite{superrad}, an amplification of the modes due to the presence of an ergoregion~\cite{zeldovich}. In this direction, however, the analogy is limited by the fact that the spatial sections of acoustic geometries are necessarily conformally flat.\footnote{In~\cite{silke} a partial mapping of the analogue metric into a Kerr black hole geometry has been done. Being the spatial part of the acoustic metric conformally flat, the mapping was possible just for the equatorial slice of the Kerr metric.}
Expanding universes can be simulated with such systems, but, for similar reasons, only de Sitter spacetime, or in general, flat ($k=0$) Friedmann--Robertson--Walker~(FRW) (see e.g.~\cite{LRR,expandinguniverse} and references therein) spacetimes can be cast in the acoustic form.
This experiment-directed work had also motivated a wealth of theoretical studies regarding the robustness of Hawking radiation against ultraviolet (UV) Lorentz symmetry breaking~\cite{HR-robust} and on its possible signatures~\cite{corr-BEC,macherparentani}.

While the above subjects have represented what we might call the main stream of research on analogue models, in more recent years there
has been growing attention on their application as toy models for emergent gravity scenarios~\cite{toymodel}. The latter are driven by the idea that gravity could be an intrinsically classical/large-scale phenomenon similar to a condensed matter state made of many fundamental constituents~\cite{BLH}. In this sense, gravity would not be a fundamental interaction but rather a large-scale/number effect, which emerges from a quite different dynamics of some elementary quantum objects. In this sense, many examples can be brought up, starting from the causal set proposal~\cite{Bombelli:1987aa}, passing to group field theory~\cite{Oriti:2007qd} or the recent quantum graphity models~\cite{Konopka:2008hp} and other approaches (see e.g.~\cite{Dreyer,Padma}).

Analogue models, and in particular Bose--Einstein condensates, have represented in this sense a very interesting test field as they are in many ways ideal systems where the mechanism of emergence of symmetries, spacetime and possibly gravitation-like dynamics, can be
comprehensively studied and understood. In particular, much attention has been devoted in the recent past to Lorentz symmetry breaking and its possible role in leading to viable mechanisms of emergent dynamics (see e.g. the related discussion in~\cite{SIGRAV}).

However,
except for the very recent work on the Abelian Higgs model~\cite{chinese}, generically the analogue models considered so far are characterized by non-relativistic fundamental equations. This would imply, in the emergent gravity perspective, a transplanckian world characterized by a preferred system of reference and a Newtonian absolute space and time. In this work we are going to consider a new analogue gravity system, namely a relativistic Bose--Einstein condensate (RBEC) for which the above issue is not present, and we shall show that also in this case an emergent spacetime structure can be found. From the point of view of emergent gravity analogues, this system can then be studied as an alternative to the standard case of  a transition from Lorentzian to Galilean relativity in the UV, showing instead a transition from Lorentz to Lorentz symmetry encoded in the different invariant speeds at different energy regimes.
Moreover,
these systems are richer than their non-relativistic counterparts as they are characterized by two kinds of propagating modes, a massless and a massive one (which disappears in the non-relativistic limit): we shall show that just the massless one is described by a generalized Klein--Gordon equation in a curved spacetime.
Finally, we will show that the analogy with gravity can be applied to RBECs also and that the resulting metric allows to produce and investigate novel classes of metrics, as for example the $k=-1$ FRW metric that we propose in the final part of the paper. Moreover the spatial slices of the resulting acoustic metric are not conformally flat, possibly permitting much wider mapping.

The plan of the paper is the following. In section~\ref{sec:bec} the relativistic description of a BEC is given. The equation describing the perturbations in such a system is described in section~\ref{sec:perturbations}, and their dispersion relation is studied in section~\ref{sec:dispersion} in different regimes. In section~\ref{sec:metric} the emergence of an acoustic metric is shown.
As an application, in section~\ref{sec:frw} a possible mapping of the relativistic metric into a FRW geometry with $k=-1$ is proposed and commented. We finally conclude with some remarks in section~\ref{sec:conclusion}, discussing what lesson can be learnt from this model.

\section{Relativistic BEC}
\label{sec:bec}

Bose--Einstein condensation, i.e. the macroscopic occupation of a single state, may occur both for relativistic and non-relativistic bosons: the main differences between their thermodynamical properties at finite temperature are due both to the different energy spectra and also to the presence, for relativistic bosons, of anti-bosons. These differences result in different conditions for the occurrence of Bose--Einstein condensation, which is possible, e.g., in two spatial dimensions for a homogeneous relativistic Bose gas, but not for its non-relativistic counterpart---and also, more importantly for our purposes, in the different structure of their excitation spectra. In this section we briefly recall the thermodynamic properties of a relativistic Bose gas~\cite{haber81}--\cite{witkowska09},
discussing how the non-relativistic limit is obtained 
and comparing with the results for non-RBECs~\cite{stringari03,pethick08}. The study of the excitation spectrum for a (generally moving and inhomogeneous) condensate is presented in the following section~\ref{sec:perturbations}.

The Lagrangian density for an interacting relativistic scalar Bose field $\hat{\phi}(\bi{x},t)$ may be written as
%
\begin{equation}
\hat{{\cal L}}=\frac{1}{c^2}\frac{\partial\hat\phi^\dagger}{\partial t}\frac{\partial\hat\phi}{\partial t}-
\bi{\nabla}\hat\phi^\dagger \cdot \bi{\nabla}\hat\phi 
-\left(\frac{m^2c^2}{\hbar^2} + V(t,\bi{x})\right) \hat{\phi}^\dagger \hat{\phi}-U(\hat\phi^\dagger\hat\phi;\lambda_i),
\label{Lagrangian}
\end{equation}
where $V(t,\bi{x})$ is an external potential depending both on time $t$ and position $\bi{x}$, $m$ is the mass of the bosons and $c$ is the velocity of light. $U$ is an interaction term and the coupling constants $\lambda_i(t,\bi x)$ can also depend on both time and position.
$U$ can be expanded as
\begin{equation}
 U(\hat\phi^\dagger\hat\phi;\lambda_i) = \frac{\lambda_2}{2}\hat{\rho}^2 + \frac{\lambda_3}{6}\hat{\rho}^3 + \cdots,
\end{equation}
where $\hat{\rho}=\hat\phi^\dagger\hat\phi$. The usual two-particle $\lambda_2 \hat{\phi}^4$ interaction corresponds to the first term $(\lambda_2/2)\hat{\rho}^2$, while the second term represents the three-particle interaction and so on.

Since the Lagrangian~\eref{Lagrangian} is invariant under the global $U(1)$ symmetry, it is possible to define the associated Noether current and conserved charge, the latter corresponding to $N-\bar{N}$, where $N$ ($\bar{N}$) is the number of bosons (anti-bosons).

Then, for an ideal homogeneous gas, when both $U$ and $V$ vanish, the relation between the chemical potential $\mu$ and the temperature $T\equiv 1/(k_B\beta)$ is given by~\cite{haber81}
\begin{equation}
N-\bar{N}=\sum_{\bi{k}} \left[ n_{\bi{k}}(\mu,\beta) - \bar{n}_{\bi{k}}(\mu,\beta)\right],
\label{mu_T}
\end{equation}
where $n_{\bi{k}}(\mu,\beta)=1/\{\exp[\beta (|E_{\bi{k}}|-\mu)]-1\}$ is the average number of bosons in the state of energy  $|E_{\bi{k}}|$, with
\begin{equation}
 E_{\bi{k}}^2= \hbar^2 k^2 c^2+m^2c^4.
\end{equation}
Similarly, $\bar{n}_{\bi{k}}(\mu,\beta)=1/\{\exp[\beta (|E_{\bi{k}}|+\mu)]-1\}$ is the corresponding number of anti-bosons.

Introducing  the number density $n=(N-\bar{N})/\Omega$ (where $\Omega$ is the volume of the system), one obtains in $d$ dimensions for non-interacting bosons the following relation between the critical temperature $T_c$ and the charge density $n$~\cite{haber81,grether07}:
\begin{equation}
n=C \int_{0}^{\infty} \rmd k k^{d-1} \frac{\sinh{(\beta_c mc^2)}}
{\cosh{(\beta_c |E_{\bi{k}}|)}-\cosh{(\beta_c mc^2)}},
\label{T_c}
\end{equation}
where $C=1/(2^{d-1} \pi^{d/2}\Gamma(d/2))$ is a numerical coefficient.

From~\eref{T_c} one can readily derive the non-relativistic and the ultra-relativistic limits: the former is obtained when $k_B T_c \ll mc^2$. In this limit the contribution of anti-bosons to~\eref{mu_T} can be neglected (so $n\approx N / \Omega$) and one obtains
\begin{equation}
k_B T_c=\frac{2\pi \hbar^2}{n} \left( \frac{n}{\zeta(d/2)} \right)^{2/d}
\label{T_c_NR}
\end{equation}
($\zeta$ denotes the Riemann zeta function), which is the usual result for the critical temperature of a non-relativistic ideal Bose gas \cite{stringari03,pethick08}.
In the ultra-relativistic limit, $k_B T_c \gg mc^2$, one obtains
\begin{equation}
\left( k_B T_c \right)^{d-1}=\frac{\hbar^d c^{d-2} \Gamma{(d/2)} (2\pi)^d}
{4 m \pi^{d/2} \Gamma(d) \zeta(d-1)}  \,n.
\label{T_c_UR}
\end{equation}
%
From now on we focus on the case $d=3$, where the homogeneous relativistic ideal Bose gas undergoes Bose-Einstein condensation.

At $T\ll T_c$, when the relativistic bosons condense, it is then possible to describe the dynamics of the condensate at the mean-field level by performing the substitution $\hat{\phi} \to \phi$: the order parameter $\phi$ satisfies then the classical equation
%
\begin{equation}
\frac{1}{c^2} \frac{\partial^2 \phi}{\partial t^2} - \bi{\nabla}^2 \phi
+\left(\frac{m^2 c^2}{\hbar^2} + V(t,\bi{x})\right)\phi
+U'(\rho;\lambda_i(t,\bi{x})) \phi=0,
\label{NL_KG}
\end{equation}
where $\rho=\phi^\ast\phi$ and $'$ denotes the derivative with respect to $\rho$. The nonlinear Klein--Gordon equation~\eref{NL_KG} gives the dynamics of the relativistic condensates, and in then non-relativistic limit the Gross--Pitaevskii equation is retrieved.

Adopting the standard definition for the box operator in flat spacetime,
\begin{equation} 
\Box=\eta^{\mu\nu}\partial_\mu\partial_\nu=-\frac{1}{c^2}\partial_t^2+\nabla^2,
\end{equation}
equation~\eref{NL_KG} can be written as
\begin{equation}\label{eq:eqphi}
\Box\phi-\left(\frac{m^2 c^2}{\hbar^2} + V\right)\phi
-U'(\rho;\lambda_i) \phi=0.
\end{equation}

\section{Analysis of perturbations}
\label{sec:perturbations}

In this section, we study the excitation spectrum of perturbations on a condensate obeying the classical wave function equation~\eref{eq:eqphi}. The field $\hat\phi$ can be written as a classical field (the condensate) plus perturbation:
\begin{equation}\label{eq:exp}
 \hat\phi = \phi(1+\hat\psi).
\end{equation}
It is now worth noting that the expansion in~\eref{eq:exp} can be linked straightforwardly to the usual expansion~\cite{LRR,stringari03,pethick08} in phase and density perturbations $\hat\theta_1$, $\hat\rho_1$:
\begin{equation}
 \frac{\hat\rho_1}{\rho}=\frac{\hat\psi+\hat\psi^\dagger}2,\quad \hat\theta_1=\frac{\hat\psi-\hat\psi^\dagger}{2\rmi}.
\end{equation}
The equation for the quantum field $\hat\psi$ describing the perturbations is
\begin{equation}\label{eq:pert}
 \Box\hat\psi+2\eta^{\mu\nu}(\partial_\mu\ln{\phi})\partial_\nu\hat\psi-\rho\, U''(\rho;\lambda_i)(\hat\psi+\hat\psi^\dagger)=0.
\end{equation}
It is now quite convenient to adopt a Madelung representation for the complex mean field $\phi$ and decompose it into two real fields, its modulus $\sqrt{\rho(x,t)}$ and its phase $\theta(x,t)$
%
\begin{equation}
 \phi=\sqrt{\rho}\,\rme^{\rmi\theta},
\end{equation}
the logarithm in~\eref{eq:pert} then becomes
\begin{equation}
 \partial_\mu\ln\phi=\frac{1}{2}\partial_\mu\ln\rho+\rmi\,\partial_\mu\theta.
\end{equation}
For convenience we define the following quantities:
\begin{eqnarray}
 u^\mu \equiv \frac{\hbar}{m}\eta^{\mu\nu}\partial_\nu\theta,\\
 c_{0}^2\equiv\frac{\hbar^2}{2m^2}\rho\, U''(\rho;\lambda_i),\label{eq:defc0}\\
 T_\rho\equiv-\frac{\hbar^2}{2m}\left(\Box+\eta^{\mu\nu}\partial_\mu\ln{\rho}\,\partial_\nu\right)=-\frac{\hbar^2}{2m\rho}\eta^{\mu\nu}\partial_\mu\rho\,\partial_\nu,
\end{eqnarray}
where the derivatives act on everything on their right,
$c_{0}$ encodes the strength of the interactions and has dimensions of a velocity and $T_\rho$ is a generalized kinetic operator that reduces, in the non-relativistic limit $c\to\infty$ and for constant $\rho$, to the standard kinetic energy operator $T$:
\begin{equation}
 T_\rho\to-\frac{\hbar^2}{2m\rho}\nabla\rho\nabla=-\frac{\hbar^2}{2m}\nabla^2=T.
\end{equation}
A straightforward physical interpretation can be given to the four-vector $u^\mu$.
One can introduce the conserved current
\begin{equation}
 j_\mu\equiv \frac{1}{2\rmi}\left(\phi\partial_\mu\phi^*-\phi^*\partial_\mu\phi\right),
\end{equation}
and show that
\begin{equation}
j_\mu=\rho\partial_\mu\theta=\rho\frac{m}{\hbar}u_\mu,
\end{equation}
hence relating $u^\mu$ to the current associated with the $U(1)$ symmetry.

In terms of these quantities and using the phase-density decomposition, the equation for the condensate classical wave function, \eref{eq:eqphi}, becomes
%
 \begin{eqnarray}
  \partial_\mu (\rho u^\mu)=0,\label{eq:cont1}\\
  -u_\mu u^\mu = c^2+\frac{\hbar^2}{m^2}\left[ V(x^\mu) + U'(\rho;\lambda_i(x^\mu))- \frac{\Box \sqrt{\rho}}{\sqrt{\rho}} \right].\label{eq:eul}
  \end{eqnarray}
%
The first equation is a continuity equation, which tells that the current $j^\mu$ defined above is conserved. The second one allows one to determine the zero-component of $u^\mu$ and, equivalently, the chemical potential, as a function of the spatial part of the fluid velocity, the strength of the interaction and the condensate density $\rho$.
%
%
%

Multiplying~\eref{eq:pert} by $\hbar^2/2m$, one can rewrite the equation for perturbations in an RBEC as
%
\begin{equation}\label{eq:dirac}
 \left[\rmi\hbar u^\mu\partial_\mu-T_\rho-mc_0^2\right]\hat\psi=mc_0\hat\psi^\dagger.
\end{equation}
%
%
Taking the Hermitian conjugate of this equation, we can eliminate $\hat\psi^\dagger$, obtaining a single equation for $\hat\psi$:
\begin{equation}
 \left[-\rmi\hbar u^\mu\partial_\mu-T_\rho-mc_0^2\right]
 \left[\rmi\hbar u^\mu\partial_\mu -T_\rho-mc_0^2\right]\hat\psi
 =m^2c_0^4\,\hat\psi,
\end{equation}
and, with some simple manipulations, we obtain
\begin{equation}\label{eq:psifluid}
 \left\{\left[\rmi\hbar u^\mu\partial_\mu+T_\rho\right]
 \frac{1}{c_0^2}
 \left[-\rmi\hbar u^\mu\partial_\mu+T_\rho\right]
 -\frac{\hbar^2}{\rho}\eta^{\mu\nu}\partial_\mu\rho\,\partial_\nu\right\}\hat\psi=0.
\end{equation}
%
This is the generalization to a relativistic condensate of the equation describing the propagation of the linearized perturbations on top of a non-relativistic BEC, in the same form as in~\cite{macherparentani}. It is worth stressing that~\eref{eq:psifluid} is implied by~\eref{eq:dirac} but the converse is not true.
Nevertheless~\eref{eq:psifluid} is enough for our aim, since we are interested only in the dispersion relation and in the metric describing the field propagation, which can be obtained from the modified Klein--Gordon equation in a curved background.

As an application of~\eref{eq:psifluid}, we consider a moving condensate with homogeneous density, $V(t,\bi{x})=0$ and $\lambda_i$ being constant both in space and time.
Let us choose
\begin{equation}
\phi(t,\bi{x})=
\phi_0 \rme^{\rmi \left( \bi{q} \cdot \bi{x} -\mu t/\hbar\right)},
\end{equation}
one has 
\begin{eqnarray}
 \mu \equiv mcu^0,\label{eq:mu}\\
 \bi{q}\equiv m \bi{u}/\hbar,\\
u^\mu \partial_\mu =(\mu/mc^2)\partial/\partial t+(\hbar/m)\bi{q} \cdot \bi{\nabla}.
\end{eqnarray}
Moreover~\eref{eq:eul} reduces to
\begin{equation}\label{eq:homeul}
\mu^2=\hbar^2 c^2 q^2+m^2c^4+\hbar^2 c^2 U'(\rho_0;\lambda_i),
\end{equation}
where $\rho_0=\phi_0 \phi_0^\ast$.

Note that $\mu$ is the relativistic chemical potential of the condensate, which is related to the non-relativistic counterpart $\mu_{\rm NR}$ via the identity $\mu=mc^2+\mu_{\rm NR}$. In particular, from~\eref{eq:homeul}, in the non-relativistic limit $\mu_{\rm NR}\ll mc^2$ and $\mu\approx mc^2$. In the general case, when the fluid is not homogeneous, the same argument implies, from~\eref{eq:eul}, that $u^0\approx c$.

Setting $\psi \propto {\exp}[\rmi \left( \bi{k} \cdot \bi{x} -\omega t\right)]$ one obtains from~\eref{eq:psifluid}
%
\begin{eqnarray}\label{eq:fourierexp_gen}
\fl
\left(- \bi{u} \cdot \bi{k}+
\frac{u^0}{c}\omega-\frac{\hbar }{2 mc^2}\omega^2+\frac{\hbar}{2m}k^2\right) 
\nonumber \\ \times
\left( \bi{u} \cdot \bi{k}
-\frac{u^0}{c}\omega-\frac{\hbar }{2 mc^2}\omega^2+\frac{\hbar}{2m}k^2\right)
 -\left(\frac{c_0}{c}\right)^2\omega^2+c_0^2 k^2=0.
\end{eqnarray}
%
For a condensate at rest ($\bi{u}=0$), the spectrum~\eref{eq:fourierexp_gen} was derived in \cite{haber81,witkowska09,andersen07}.

The non-relativistic limit of~\eref{eq:psifluid} can be recovered by letting $c\to\infty$. This implies $\mu=mc^2+\mu_{\rm NR}\approx mc^2$ and, equivalently $u^0\approx c$. In this limit, the mode equation becomes
%
\begin{equation}\label{eq:psiNR}
\fl
 \left\{\left[\rmi\hbar\left(\partial_t+\ub\cdot\nabla\right)+T_{\rho\,{\rm NR}}\right]
 \frac{1}{c_0^2}
 \left[-\rmi\hbar\left(\partial_t+\ub\cdot\nabla\right)+T_{\rho\,{\rm NR}}\right]
 -\frac{\hbar^2}{\rho}\nabla\rho\,\nabla\right\}\hat\psi=0,
\end{equation}
where
\begin{equation}
 T_{\rho\,{\rm NR}}\equiv-\frac{\hbar^2}{2m\rho}\nabla\rho\nabla.
\end{equation}
\Eref{eq:psiNR} reduces to the usual equation describing the propagation of perturbations in a non-relativistic BEC.

\section{The dispersion relation}
\label{sec:dispersion}
%
Before showing how the propagation of phonons in an RBEC can be described through an effective metric, it is worthwhile to analyze the dispersion relation of such perturbations. To do that
we assume in this section that $u$, $\mu$, $\rho$ and $c_0$ are constant both in space and in time. For simplicity we start to analyze the case of background fluid at rest, $\ub=0$. As discussed after~\eref{eq:psifluid}, in order to study the dispersion relation, one can directly work on~\eref{eq:psifluid}, which becomes in this case
%
%
%
\begin{equation}
 \left[\left(\rmi\frac{u^0}{c}\partial_t-\frac{\hbar}{2m}\Box\right)
 \left(-\rmi\frac{u^0}{c}\partial_t-\frac{\hbar}{2m}\Box\right)
 -c_0^2\,\Box\right]\hat\psi=0,
\end{equation}
which can be rewritten expanding the $\Box$ operator as
\begin{eqnarray}\label{eq:flat}
\fl
 \left[\left(\rmi\frac{u^0}{c}\partial_t+\frac{\hbar }{2 mc^2}\partial_t^2-\frac{\hbar\nabla^2}{2m}\right)
  \right.\nonumber\\ \left.\times
 \left(-\rmi\frac{u^0}{c}\partial_t+\frac{\hbar }{2 mc^2}\partial_t^2-\frac{\hbar\nabla^2}{2m}\right)
 +\left(\frac{c_0}{c}\right)^2\partial_t^2-c_0^2\,\nabla^2\right]\hat\psi=0~.
\end{eqnarray}
The previous equation can be solved exactly by Fourier modes $\exp(-\rmi\omega t+\rmi \bi{k}\cdot \bi{x})$
\begin{eqnarray}\label{eq:fourierexp}
\fl 
\left(\frac{u^0}{c}\omega-\frac{\hbar }{2 mc^2}\omega^2+\frac{\hbar}{2m}k^2\right)
 \left(-\frac{u^0}{c}\omega-\frac{\hbar }{2 mc^2}\omega^2+\frac{\hbar}{2m}k^2\right)
 -\left(\frac{c_0}{c}\right)^2\omega^2+c_0^2 k^2=0,
\end{eqnarray}
whose solution is:
\begin{eqnarray}\label{eq:dispersion}
\fl
 \omega^2_\pm=c^2
 \left\{k^2+2\left(\frac{mu^0}{\hbar}\right)^2 \left[1+\left(\frac{c_0}{u^0}\right)^2\right]
\right.\nonumber\\ \left. 
 \pm 2\left(\frac{mu^0}{\hbar}\right)
 \sqrt{k^2+\left(\frac{mu^0}{\hbar}\right)^2\left[1+\left(\frac{c_0}{u^0}\right)^2\right]^2}
 \right\}.
\end{eqnarray}
%

The above equation represents the dispersion relation for the modes in an RBEC and it is the generalization of the non-relativistic Bogoliubov dispersion relation.

\begin{figure}
\begin{flushright}
\includegraphics[width=.85\columnwidth]{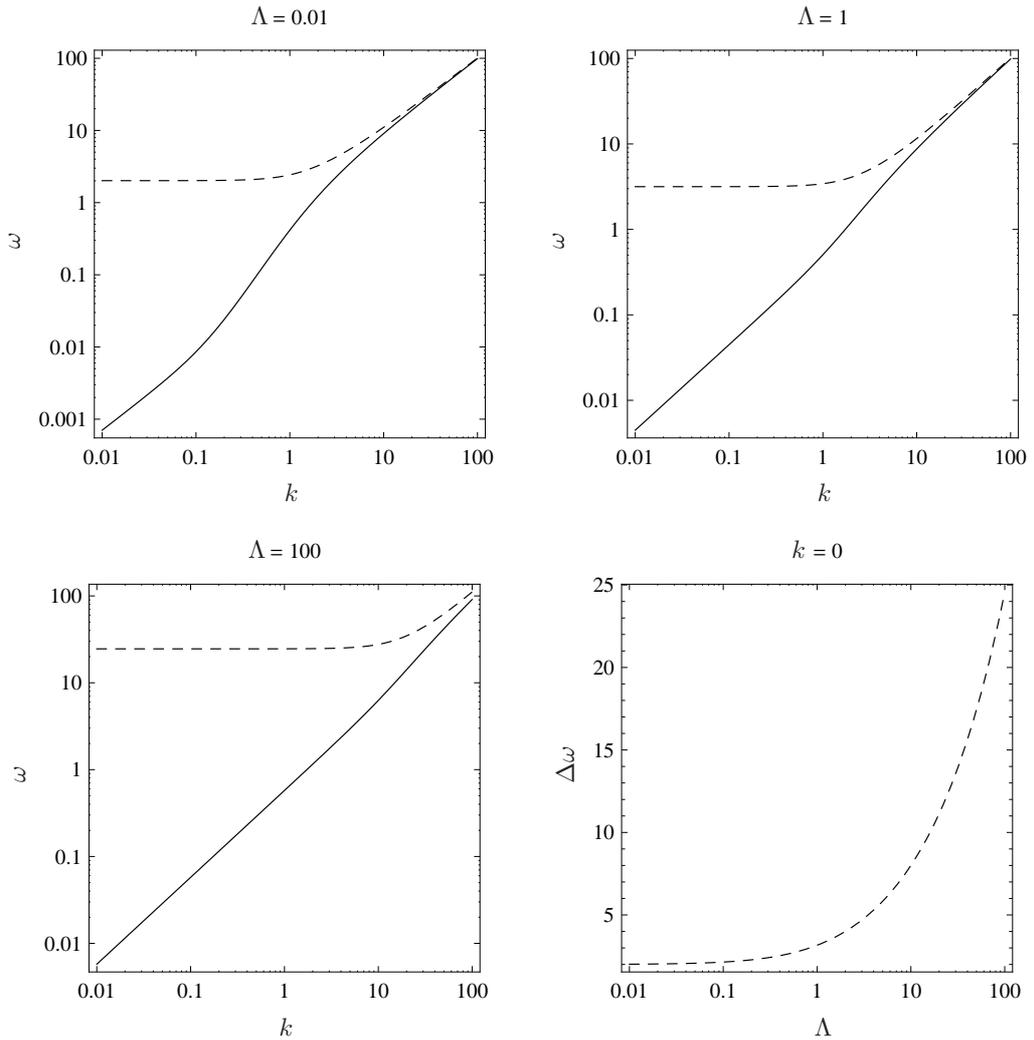}
\end{flushright}
\caption{Excitation spectrum~\eref{eq:dispersion} for a condensate at rest ($\bi{u}=0$) with interaction $U(\rho)=\lambda_2 \rho^2/2$ and $\Lambda = (\hbar/mc)^2\lambda_2\rho_0 =0.01,\,1,\,100$. Solid line: gapless branch $\omega_-$, dashed line: gapped branch $\omega_+$. $\omega$ in units of $mc^2/\hbar$, $k$ in units of $mc/\hbar$. Bottom right: 
plot of the the gap $\Delta \omega=\omega_+-\omega_-$ at $k=0$, which in these units is given by $\Delta \omega(k=0)=\sqrt{4+3\Lambda}$.}
\label{fig:spectrum_at_rest}
\end{figure}

One can see from~\eref{eq:dispersion} that $\omega^2_\pm \geqslant0$, i.e. there is no dynamical instability: in figure~\ref{fig:spectrum_at_rest} we plot the dispersion relation~\eref{eq:dispersion} for the interaction $U(\rho)=\lambda_2 \rho^2/2$ for different values of the dimensionless parameter $\Lambda \equiv  (\hbar/mc)^2\lambda_2\rho_0$. In figure~\ref{fig:spectrum_at_rest}, we also plot the gap at $k=0$ between $\omega_+(k=0)$ and $\omega_-(k=0)$.

One sees from (\ref{eq:dispersion}) that the different regimes allowed for the excitations of the systems are determined by the relative strength of the the first two terms on the right-hand side of (\ref{eq:dispersion}) (note that the same terms enter in the square root).
Hence, in what follows we shall analyze the RBEC excitations dispersion relation~\eref{eq:dispersion} in some significant limits. The results are summarized at the end of this section in table~\ref{tab:regimes}.

\subsection{Low momentum limit}
\label{sec:low}

We begin our study by looking at
the low momentum limit of the dispersion relation (\ref{eq:dispersion}) characterized as
\begin{equation}\label{eq:lowkcond}
|k| \ll \frac{m u^0}{\hbar }\left[1+\left(\frac{c_0}{u^0}\right)^2\right] \equiv  \frac{m u^0}{\hbar }(1+b),
\end{equation}
where, for later convenience, we have introduced the dimensionless parameter
\begin{equation}
b\equiv \left( \frac{c_0}{u^0} \right)^2.
\label{eq:b}
\end{equation}

Under the above assumption (\ref{eq:lowkcond}), the square root in~\eref{eq:dispersion} can be expanded as
\begin{eqnarray}\label{eq:lowk}
\fl
 \omega^2_\pm\approx c^2\!
 \left[k^2+2\left(\frac{m u^0}{\hbar }\right)^2\!\!(1+b)
 \pm 2\left(\frac{m u^0}{\hbar }\right)^2\!\!(1+b)
 \pm \frac{k^2}{1+b}
 \mp \frac{k^4}{4(mu^0/\hbar)^2(1+b)^3}
 \right].
\end{eqnarray}
%



\subsubsection{Gapless excitations.}
Let us focus for the moment on the branch corresponding to the lower sign in the above dispersion relation. One obtains
\begin{equation}\label{eq:lowkm}
 \omega_{-}^2\approx c^2
 \left[\frac{b}{1+b}k^2
 + \frac{k^4}{4(mu^0/\hbar )^2(1+b)^3}
 \right].
\end{equation}
This is the dispersion relation of a massless quasi-particle and has the same form of the Bogoliubov dispersion relation in a non-relativistic BEC. 
Let us study now when the $k^2$ term dominates over the $k^4$ one and vice versa and check if the conditions obtained for these two regimes are compatible with~\eref{eq:lowkcond}.
\paragraph{Phononic (IR relativistic) regime:} It is easy to see that the quartic term $k^4$ can be neglected whenever
\begin{equation}
 |k|\ll 2\frac{m c_0}{\hbar }(1+b).
\end{equation}
This condition is always compatible with~\eref{eq:lowkcond}. Depending on the value of $b$, they can be written in the compact form
\begin{equation}\label{eq:lowlowk}
 |k|\ll \frac{m u^0}{\hbar }(1+b)\min\{1,2\frac{c_0}{u^0}\}.
\end{equation}

In this limit the dispersion relation reduces to the usual phonon dispersion relation
\begin{equation}\label{eq:lineardisp}
 \omega^2_{-}=c_s^2 k^2,
\end{equation}
where the speed of sound $c_s$ is defined as
\begin{equation}\label{eq:cs}
 c_s^2\equiv\frac{(c\, c_0/u^0)^2}{1+(c_0/u^0)^2}=\frac{c^2 b}{1+b}.
\end{equation}
%
%

\paragraph{Newtonian (UV Galilean) regime:} The opposite regime, in which the $k^2$ can be neglected with respect to $k^4$, is defined by the following condition:
\begin{equation}\label{eq:k2negl}
 |k|\gg \left(\frac{mc_0}{\hbar }\right)(1+b).
\end{equation}
This regime is present only in the low-coupling case when $b\ll 1$, since only in this case~\eref{eq:k2negl} and~\eref{eq:lowkcond} are compatible. They can be summarized as
\begin{equation}
 \frac{mc_0}{\hbar}\ll |k|\ll \frac{m u^0}{\hbar },
\end{equation}
Under these assumptions the dispersion relation becomes
\begin{equation}\label{eq:nonreldisp}
 \hbar\omega_{-}=\frac{(\hbar k)^2}{2mu^0/c}=\frac{(\hbar k)^2}{2\mu/c^2},
\end{equation}
which represents the Newtonian dispersion relation for massive particles with effective mass $mu^0/c=\mu/c^2$. This is not surprising, since it is known from non-relativistic BEC that decreasing the wavelength of the perturbation, the atomic structure of the condensate emerges.

In this case (an RBEC) we see that, even if the structure of the Lagrangian is fully relativistic, when the energy of the perturbations is not much higher than the chemical potential, the bosons are moving with non-relativistic speed, and the dispersion relation is Newtonian. Noticeably, the mass of the non-relativistic quasi-particles is not the mass of the bosons $m$, but it is an effective mass $\mu/c^2$, the rest mass corresponding to an energy equal to the chemical potential $\mu$. This is not surprising since the chemical potential actually represents the energy needed to add a particle to the condensate.
%

\paragraph{The non-relativistic limit:} We discuss the non-relativistic limit of the gapless dispersion relation~\eref{eq:lowk} starting from the behavior of the parameter $b=(c_0/u^0)^2$. As shown in section~\ref{sec:perturbations}, $u^0\to c$ in this limit when the fluid flows with a velocity much smaller than the speed of light. Moreover, in the same limit, the interaction must be weak, namely $c_0\ll c$, such that
\begin{equation}\label{eq:bnonrel}
 b\approx(c_0/c)^2\ll1.
\end{equation}
%
In this sense we can say that $b$ measures the relativistic nature of the condensate, taking into account both the strength of the interaction and the velocity of the fluid, as it is possible to see from the definition of $b$~\eref{eq:b}, using~\eref{eq:defc0}, \eref{eq:mu} and~\eref{eq:homeul}.
Using~\eref{eq:bnonrel} in~\eref{eq:lowk} it is easy to obtain the Bogoliubov dispersion relation for a non-relativistic BEC (see, for example,~\cite{LRR})
\begin{equation}\label{eq:lowkmnonrel}
  \omega_{-}^2\approx c_0^2
 \left[k^2
 + \frac{k^4}{(2mc_0/\hbar )^2}
 \right].
\end{equation}
It is worth noting that $c_s$~\eref{eq:cs} is, in general, different from the speed of sound $c_0$ in a non-relativistic BEC, but, as it is evident from the above equation, $c_s$ reduces correctly to $c_0$ in the non-relativistic limit. 
The functional form of the dispersion relation~\eref{eq:lowkm} is unchanged with respect to~\eref{eq:lowkmnonrel}. Therefore, at this energy regime, there is no hint of the relativistic nature of the underlying BEC, without knowing the microphysics governing the elementary constituents. In other words, at this energy scale, a phononic observer cannot be aware of the relativistic nature of the condensate.

%
Finally, let us also note that in this limit, condition~\eref{eq:lowlowk}, defining the phononic regime, reduces to the usual
\begin{equation}	
 |k|\ll 2\frac{mc_0}{\hbar},
\end{equation}
where $mc_0/\hbar$ is the inverse of the healing length. When this condition is satisfied, one obtains the phononic dispersion relation
\begin{equation}
 \omega_-^2 = c_0^2 k^2,
\end{equation}
while, in the opposite regime, the dispersion relation describes Newtonian particles with mass $m$
\begin{equation}
 \hbar\omega_-=\frac{\hbar^2k^2}{2m},
\end{equation}
as expected taking directly the non-relativistic limit of~\eref{eq:nonreldisp} and putting $\mu\approx m c^2$.

\subsubsection{Gapped excitations.}

A very different situation appears for the upper sign solution in~\eref{eq:lowk}:
\begin{equation}\label{eq:lowkgap}
 \omega^2_+\approx c^2
 \left[4\left(\frac{m u^0}{\hbar }\right)^2(1+b)
 + \frac{2+b}{1+b}k^2
 - \frac{k^4}{4(m u^0/\hbar )^2(1+b)^3}
 \right].
\end{equation}
The only regime allowed is now the one in which the $k^2$ term dominates over the $k^4$ term.\footnote{In fact, the latter term would dominate if
%
 $k^2\gg4(mu^0/\hbar)^2(1+b)^2(2+b)>8(mu^0/\hbar)^2(1+b)^2$,
%
but the above condition is clearly not satisfied when~\eref{eq:lowkcond} holds.}
In this limit~\eref{eq:lowkgap} becomes
\begin{equation}\label{eq:dispgap}
 \omega^2_+=\frac{m_{\rm eff}^2\,c_{s,{\rm gap}}^4}{\hbar^2}+c_{s,{\rm gap}}^2 k^2,
\end{equation}
where we have defined
\begin{eqnarray}
 c_{s,{\rm gap}}^2 \equiv \frac{2+b}{1+b}\,c^2~,\quad\\
 m_{\rm eff} \equiv 2\frac{(1+b)^{3/2}}{2+b}\frac{u^0}{c}m=2\frac{(1+b)^{3/2}}{2+b}\frac{\mu}{c^2}.
\end{eqnarray}
\Eref{eq:dispgap} is the dispersion relation for a massive relativistic particle of effective mass $m_{\rm eff}$ and limit speed $c_{s,{\rm gap}}$. Note that, when~\eref{eq:lowkcond} holds, and for $b\ll1$, the $k^2$ term is always dominated by the mass term. That is, this mode represents non-relativistic particles with mass $m_{\rm eff}$. When instead $b\gg1$, the $k^2$ term can be of the same order of magnitude as the mass term, and the fully relativistic dispersion relation with mass gap becomes important. The $k^2$ term dominates when
\begin{equation}
 \frac{mu^0}{\hbar }\frac{2(1+b)}{\sqrt{2+b}} \ll |k|\ll \frac{mu^0}{\hbar }(1+b).
\end{equation}

Let us now spend a few words about the physical meaning of this massive mode. Let us assume, for simplicity, that $b$ can be neglected. In this case, from the above formulae, the gap is $\Delta\omega\approx 2 mc^2/\hbar$. This is an excitation of completely different nature with respect to previously discussed phonons. The mass gap of $2m$ indicates that the lowest possible excitation in this mode requires us to create a boson--anti-boson couple. When $k\neq 0$ the massive excitation propagates: this can be seen as the creation of a couple of particles, then its annihilation, while its energy is used to create another couple close to the former one and so on. Note that, apparently, this mode can make energy travel faster than light. However, one should consider that the dispersion relation~\eref{eq:dispgap} is valid only for sufficiently low $k$, satisfying~\eref{eq:lowkcond}. It is easy to check indeed that under this condition for $k$, the group velocity $\partial\omega/\partial k$ always remains smaller than $c$.

The non-relativistic limit of the gapped dispersion relation~\eref{eq:lowkgap} is trivial. The zero-order term in $k^2$ becomes now $4m^2c^4/\hbar^2$, which diverges when taking the limit $c\to\infty$. In fact the gapped branch disappears because the energy needed to excite this mode is much larger (indeed infinite in the limit) than the typical energy scales in non-relativistic configurations. In other words, this mode cannot be excited because one cannot create particle--anti-particle couples in the non-relativistic limit.

\subsection{High momentum}
\label{sec:high}

The situation for high momentum is much simpler. When
\begin{equation}
 |k| \gg \frac{mu^0}{\hbar }\left[1+\left(\frac{c_0}{u^0}\right)^2\right]=\frac{mu^0}{\hbar }(1+b),
\end{equation}
it is easily found that the dispersion relation~\eref{eq:dispersion} assumes the very simple form
\begin{equation}
 \omega_\pm^2=c^2k^2.
\end{equation}
%
%
This is the standard relativistic dispersion relation of a massless field propagating at the speed of light $c$.
This means that, when the energy of the perturbation is much larger than the chemical potential, the Newtonian particle regime of \eref{eq:nonreldisp} is overcome. The dispersion relation assumes a novel form that was absent in the non-relativistic case. Therefore at very high momenta, the perturbations can probe the relativistic nature of the background condensate.

\fulltable
{\label{tab:regimes}Dispersion relation of gapless and gapped modes in different energy regimes. $c_s^2 = c^2 b/(1+b)$, $c_{s,{\rm gap}}^2 = c^2(2+b)/(1+b)$, $m_{\rm eff} = 2(\mu/c^2)(1+b)^{3/2}/(2+b)$, $b=(c_0/u^0)^2$.}
\scriptsize
\begin{tabular}{@{}ccccc}
  \br
  						&						&			\centre{2}{Gapless}					& Gapped\\
 \ns
						&						&				\crule{2}					&	\\
						&						&	$b\ll1$			&	$b\gg1$					&	\\
  \mr

   \multirow{2}{*}{$|k|\ll mu^0(1+b)/\hbar$}	& 	$|k|\ll2mc^0/\hbar$			&	$\omega^2 = c_s^2 k^2$	&	\multirow{2}{*}{$\omega^2 = c_s^2 k^2$}	&	\multirow{2}{*}{$\omega^2=m_{\rm eff}^2\,c_{s,{\rm gap}}^4/\hbar^2+c_{s,{\rm gap}}^2 k^2 $}	\\
  						&	$2mc^0/\hbar\ll|k|\ll mu^0/\hbar$	&	$\hbar\omega = (\hbar k)^2/2(\mu/c^2)$	&				&	\\
  \mr
  $|k|\gg mu^0(1+b)/\hbar$ 			&						&	 \multicolumn{3}{c}{$\omega^2 = c^2 k^2$}					\\
 \br
 \end{tabular}
\normalsize
\endfulltable

%

\section{The acoustic metric}
\label{sec:metric}

The propagation of phonons can be described in the formalism of quantum field theory in a curved background only when the relativistic quantum potential $T_\rho$ can be neglected in~\eref{eq:psifluid}.%
\footnote{As discussed after~\eref{eq:psifluid}, to read off the dispersion relation and the metric describing the field propagation it is enough to start from~\eref{eq:psifluid} instead of the full~\eref{eq:dirac}.}
%
This is the relativistic generalization of neglecting the quantum pressure in the Gross--Pitaevskii equation in a non-relativistic BEC.

This is possible under two conditions. Firstly, the frequency and wave-number on the gapless branch $\omega_-$, \eref{eq:lowkm}, must be sufficiently low (see~\eref{eq:lowlowk}). Secondly, one has to perform an eikonal approximation, i.e. one assumes that all the background quantities vary slowly in space and time on scales comparable with the wavelength and the period of the perturbation, respectively. This latter condition is equivalent to requiring that
\begin{equation}
 \left|\frac{\partial_t \rho}{\rho}\right|\ll\omega,\quad \left|\frac{\partial_t c_0}{c_0}\right|\ll\omega,\quad\left|\frac{\partial_t{ u_\mu}}{u_\mu}\right|\ll\omega,
\end{equation}
and the corresponding relations for the variation in space.
Under these assumptions, \eref{eq:psifluid} simplifies to
\begin{equation}
 \left[u^\mu\partial_\mu
 \frac{1}{c_0^2}
 u^\nu\partial_\nu
 -\frac{1}{\rho}\eta^{\mu\nu}\partial_\mu\rho\,\partial_\nu\right]\hat\psi=0,
\end{equation}
where, again, the differential operators act on everything on their right.
%
%
In order to recover the acoustic metric, we need to cast the above equation in the form
\begin{equation}\label{eq:metricform}
 \partial_\mu \left(f^{\mu\nu}(x,t)\partial_\nu\hat\psi\right)=0,
\end{equation}
where $f^{\mu\nu}$ is the metric density
\begin{equation}
 f^{\mu\nu}=\sqrt{-g}g^{\mu\nu}.
\end{equation}
%
To this aim, as in the non-relativistic case, one can use the continuity equation~\eref{eq:cont1}
\begin{equation}\label{eq:contdef}
 \eta^{\mu\nu}\partial_{\mu}j_\nu=\partial_\mu(\rho u^\mu)=0,
\end{equation}
to commute $\rho u^\mu$ with $\partial_\mu$
%
%
%
%
\begin{equation}\label{eq:metricpsi2}
 \partial_\mu\left[
 \frac{\rho}{c_0^2}
 u^\mu u^\nu
 -\rho\eta^{\mu\nu}\right]\partial_\nu\hat\psi=0,
\end{equation}
%
%
from which the metric density is easily read
\begin{equation}
 f^{\mu\nu}=\frac{\rho}{c_0^2}
 \pmatrix{
 -c_0^2-(u^0)^2		&	-u^0\ub				\cr
 -u^0\ub		&	c_0^2\mathbbm{1}-\ub\otimes\ub
 }~.
\end{equation}
Finally, the metric describing the propagation of phonons in an RBEC is
\begin{equation}\label{eq:becmetric}
 g_{\mu\nu}=\frac{\rho}{\sqrt{1-u_\sigma u^\sigma/c_o^2}}\left[\eta_{\mu\nu}\left(1-\frac{u_\sigma u^\sigma}{c_0^2}\right)+\frac{u_\mu u_\nu}{c_0^2}\right],
\end{equation}
in coordinates $x^\mu=(x^0,{\bi x})=(ct,{\bi x})$.

The acoustic metric for perturbations in a relativistic, barotropic and irrotational fluid flow has been very recently derived in~\cite{Molina-Visser}. The above metric can be put in the same form as in~\cite{Molina-Visser} by a few variable redefinitions. To achieve this aim, let us define a four-velocity $v^\mu$:
\begin{equation}
 v^\mu \equiv \frac{c}{\|u\|}u^\mu,\quad \|u\| \equiv \sqrt{-u_\sigma u^\sigma}.
\end{equation}
With this definition, it is possible to generalize~\eref{eq:cs} when the spatial part of the four-vector is different from zero. To keep all the formulas in a covariant form, the scalar speed of sound $c_s$ must be defined as
\begin{equation}\label{eq:scalarcs}
 c_s^2 = \frac{c^2\, c_0^2/\|u\|^2}{1+c_0^2/\|u\|^2}.
\end{equation}
Using the above definitions, the metric~\eref{eq:becmetric} reduces to
\begin{equation}\label{eq:mattmetric}
 g_{\mu\nu}=\rho\frac{c}{c_s}\left[\eta_{\mu\nu}+\left(1-\frac{c_s^2}{c^2}\right) \frac{v_\mu v_\nu}{c^2}\right].
\end{equation}
%
This metric is manifestly conformal to that found in~\cite{Molina-Visser}. However, it is worth checking that they are exactly the same metric, comparing the conformal factors too. We have to show that $\rho c/c_s$ matches perfectly with the $\tilde n_0^2c^2/\tilde c_s(\tilde\rho_0+\tilde p_0)$ in~\cite{Molina-Visser}. All the quantities of~\cite{Molina-Visser} will be tilded to avoid confusion with ours.
From (54) of~\cite{Molina-Visser} we can write
\begin{equation}
 ||u|| = C\frac{\tilde\rho_0+\tilde p_0}{\tilde n_0},
\end{equation}
where $C$ is a constant and, by definition, $||u||=\nabla\tilde\Theta$. Note that from (46) of~\cite{Molina-Visser} $\tilde n_0 = ||u||\rho$, such that, apart from irrelevant constant factors,
\begin{equation}
 \frac{\tilde n_0^2}{\tilde c_s(\tilde\rho_0+\tilde p_0)} \propto \frac{\rho}{c_s},
\end{equation}
and also the conformal factors coincide.

\section{An application: $k=-1$ FRW metric}
\label{sec:frw}

Mapping the metrics describing expanding universes is one of the most interesting applications of analogue models.
However, only a mapping of the de Sitter spacetime and the $k=0$ FRW has been obtained so far \cite{LRR,expandinguniverse}.%
\footnote{%
Since every FRW metric is conformally flat, the natural attempt to reproduce such a spacetime within the framework of the standard Eulerian fluid-gravitational analogy is to set $v=0$ and $c_s={\rm constant}$.
However, with this choice only stationary universes can be reproduced, because the acoustic conformal factor $\rho$ cannot depend on time by the continuity equation. Of course, one may try to map a non-flat FRW metric onto an acoustic one by allowing for time-dependent $v$, $c_s$ and $\rho$.
Nevertheless, this task (even though not forbidden by any no-go theorem) would be much more difficult than the natural procedure obtained by using a relativistic fluid.
}%

Here we show how the above acoustic metric can be mapped into the FWR metric with $k=-1$. This is just an application, but it is important to stress how the category of metrics that one can map in this acoustic form has been immediately generalized.

We start by rewriting~\eref{eq:mattmetric} in spherical coordinates, assuming isotropy and $v^\theta=v^\phi=0$:
\begin{equation}
g_{\mu\nu}=\rho\frac{c}{c_s}
\pmatrix{
 -1+A(v^t)^2/c^2  & - A v^tv^r/c^2  & 0 & 0 \cr
 - Av^tv^r/c^2   & 1 + A(v^r)^2/c^2 & 0 & 0 \cr
 0 & 0 & r^2 & 0 \cr
 0 & 0 & 0 & r^2 \sin ^2(\theta )
 }.
\end{equation}
where
\begin{equation}
 A\equiv 1-\frac{c_s^2}{c^2}.
\end{equation}
Defining the new coordinate $\tau$ and $\xi$ as
\begin{equation}
\cases{
 \tau = \sqrt{c^2 t^2-r^2},  \cr
 \xi =  \frac{r}{\sqrt{c^2 t^2-r^2}}
}
\end{equation}
and choosing the following velocity profile:
\begin{equation}\label{eq:vfrw}
\eqalign{
 v^t = c\,\sqrt{1+\xi^2},\\
 v^r = c\,\xi,
}
\end{equation}
one can write the inverse transformations as
\begin{equation}
\cases{
 t = \tau \frac{\sqrt{1+\xi^2}}{c} = \tau \frac{v^t}{c}, \\
 r =  \tau \xi = \tau \frac{v^r}{c}.
 }
\end{equation}

To investigate the physical meaning of these coordinates, it is worth applying the above transformation also to the underlying Minkowski spacetime, seen by the condensate. The fluid velocity $v$ can be transformed to
\begin{equation}\label{eq:vfrwtauxi}
\eqalign{
 v^\tau = c,\\
 v^\xi = v^\theta = v^\phi = 0.
}
\end{equation}
Therefore they represent coordinates comoving with the fluid. The flat Minkowski line element $\rmd s_{\rm M}$ is
\begin{equation}\label{eq:minkowski}
 \rmd s_{\rm M}^2 = -c^2\rmd t^2+\rmd r^2+r^2 \rmd\Omega^2
  = -\rmd\tau^2+\tau^2\left(\frac{\rmd\xi^2}{1+\xi^2}+\xi^2 \rmd\Omega^2 \right),
\end{equation}

Going back to the acoustic metric, with the above choices the acoustic line element $\rmd s^2$ becomes
\begin{equation}\label{eq:lineelement}
 \rmd s^2=\rho\frac{c}{c_s}\left[-\frac{c_s^2}{c^2}\rmd\tau^2+ \tau ^2\left(\frac{\rmd\xi^2}{1+\xi^2}+ \xi^2\rmd\Omega^2\right)\right].
\end{equation}
One may notice that, when both $c_s$ and $\rho$ depend only on $\tau$, the above expression represents the line element of a FRW spacetime with hyperbolic space sections:
\begin{equation}
 \rmd s^2=-\rho(\tau)\frac{c_s(\tau)}{c}\rmd\tau^2
 +\tau^2 \rho(\tau)\frac{c}{c_s(\tau)}\left(\frac{\rmd\xi^2}{1+\xi^2}+ \xi^2\rmd\Omega^2\right).
\end{equation}

However $\rho$ cannot be chosen freely because it is related through the continuity equation~\eref{eq:cont1} to the velocity $v^\mu = u^\mu / ||u||$, which has already been fixed in~\eref{eq:vfrw}. Nonetheless, it is easy to show that assuming $\rho=\rho_0$ constant both in $\tau$ and $\xi$, one can satisfy the continuity equation with an appropriate $||u||$. Using~\eref{eq:vfrwtauxi} and~\eref{eq:minkowski} the continuity equation simplifies to
\begin{equation}
 \partial_\tau(\tau^3||u||)=0,
\end{equation}
from which
\begin{equation}
 ||u|| = c\frac{\Xi(\xi)}{\tau^3},
\end{equation}
where $\Xi$ is an arbitrary function.

It is therefore possible to define a new time coordinate
\begin{equation}\label{eq:timeT}
\rmd T \equiv \sqrt{\rho_0\frac{c_s(\tau)}{c}}\rmd\tau
\end{equation}
and a function $a(T)$
\begin{equation}\label{eq:Ttau}
 a(T) \equiv \tau \sqrt{\rho_0\frac{c}{c_s(\tau)}},
\end{equation}
where $\tau(T)$ is implicitly defined in~\eref{eq:timeT}. In this way, the line element assumes the familiar $k=-1$ FRW form
\begin{equation}\label{eq:frw}
 \rmd s^2=-\rmd T^2+ a(T)^2\left(\frac{\rmd\xi^2}{1+\xi^2}+ \xi^2\rmd\Omega^2\right).
\end{equation}
At least from a mathematical standpoint, one has the freedom to choose an arbitrary form for $a(T)$. This is equivalent to choosing $c_s$ as an arbitrary function of $\tau$. Using~\eref{eq:timeT} and~\eref{eq:Ttau}
\begin{equation}
 \tau(T)=\sqrt{\frac{2}{\rho_0}\int_0^T{a(T'})dT'},
\end{equation}
from which one can get $T$ as a function of $\tau$ given a generic $a(T)$, and $c_s(\tau)$ is
\begin{equation}
 c_s(\tau)=\frac{\rho_0c\tau^2}{a(T(\tau))^2}.
\end{equation}
Fixing $c_s(\tau)$ means fixing $U(\rho_0;\lambda_i(x^\mu))$ in~\eref{eq:defc0}, and hence $\lambda_i$, as given functions of $x^\mu$. Indeed, from~\eref{eq:scalarcs}, $c_0(x^\mu)$ depends both on $c_s(\tau)$ and $||u(x^\mu)||$, the latter given by the continuity equation, as discussed above. Finally, going back to the Euler equation~\eref{eq:eul}, for constant $\rho$, one can determine also $V(x^\mu)$, because all the other functions $||u||$ and $U$ have already been fixed. Using a two-body interaction $U(\rho;\lambda_i)=\lambda_2(x^\mu)\rho^2/2$ one finds
\begin{eqnarray}
 \lambda_2(\tau,\xi)	= \frac{2m^2c^2}{\rho_0\hbar^2}\frac{c_s(\tau)^2}{c^2-c_s^2(\tau)}\left(\frac{\Xi(\xi)}{\tau^3}\right)^2,\\
 V(\tau,\xi)		= -\frac{m^2c^2}{\hbar^2}\left[1+\frac{c^2+c_s^2(\tau)}{c^2-c_s^2(\tau)}\left(\frac{\Xi(\xi)}{\tau^3}\right)^2\right].
\end{eqnarray}
Note that for a given FRW spacetime and hence given $a(T)$ and $c_s(\tau)$, one still has the freedom to choose the most convenient $\Xi(\xi)$.

\section{Conclusions}
\label{sec:conclusion}

In this paper, we have applied for the first time the analogy between gravity and condensed matter to an RBEC. Starting from the relativistic description of Bose--Einstein condensates, we have studied the propagation of excitations on top of it in a very generic framework, allowing for non-homogeneous density and nontrivial flow velocity.
The full dispersion relation for the modes is given and analyzed in different regimes (see table~\ref{tab:regimes}): there are two branches,
a gapless one and one with mass gap. The former one has three different regimes: for low momenta (wavelength larger than the healing length in the non-relativistic limit) the dispersion relation is linear $\omega^2 = c_s^2 k^2$. Then, for intermediate energies, lower than the chemical potential of the condensate, the dispersion relation becomes that of a Newtonian massive particle, just as for the non-RBEC case, when the wavelength is larger than or comparable with the healing length. Finally, for an RBEC, a third regime appears when the energy of the perturbation is larger than the chemical potential, and the dispersion relation becomes linear again ($\omega^2= c^2 k^2$), but with velocity equal to the speed of light.
The second branch has instead a mass gap, showing two different regimes. In this case, the dispersion relation describes, for low momenta, a relativistic massive particle with some effective mass and an effective limit speed apparently larger than the speed of light. However, this limit speed cannot be reached by such perturbations because at energy exceeding the chemical potential, the dispersion relation enters the second regime, in which the frequency is linearly proportional to the speed of light $c$. In the non-relativistic limit this branch is inaccessible because the energy needed to excite such a mode diverges in this limit. All these features are evident from the plot of the two branches in figure~\ref{fig:spectrum_at_rest}.

For sufficiently low momenta (a generalization of the non-relativistic condition involving the wavelength of the perturbations and the healing length), it is possible to describe the propagation of the phononic mode (massless branch) through the gravitational analogy, and the corresponding acoustic metric is given. This metric allows for a wider range of applications than the usual one obtained in non-relativistic fluids. We propose as an example the mapping to the $k=-1$ FRW metric for expanding universes. We also saw how it can be used to also map spacetimes whose spatial section is not conformally flat. This could give the possibility to map even more complicated metrics (such as for example the Kerr metric)  or to build novel structures not allowed in the non-relativistic case.

This application has another important feature. Differently from all the previous systems where the analogy with gravity has been applied, the system investigated here is fully relativistic. Therefore, it does not suffer from the troubles related to the breaking of the Lorentz symmetry due to the emergence at sufficiently short wavelengths of the Newtonian structure of the underlying fluid and of its constituents. Hence, the RBEC is the first example of emergent Lorentzian spacetime from a Lorentzian background, showing therefore a Lorentz-to-Lorentz symmetry transition at high frequencies. Transitions from an IR Lorentz symmetry to a UV Galilean one are known to be severely constrained (see e.g.~\cite{Mattingly}--\cite{MacLib}). Furthermore, in effective field theories they do tend to `percolate' from the UV to the IR via renormalization group effects, by showing up in non-suppressed corrections to the low energy propagators of elementary particles \cite{Collins,IRS}. We do wonder if situations like the one just exemplified by this novel analogue model of gravity might show a less severe behavior and ameliorate this naturalness problem common to most of UV Lorentz breaking theories. We hope to investigate this issue in future works.

\ack
\addcontentsline{toc}{section}{Acknowledgments}
We thank L Sindoni, M Visser and S Weinfurtner for useful discussions and comments. S Fagnocchi, MK and AT have been supported by the grants INSTANS (from ESF) and 2007JHLPEZ (from MIUR).


\section*{References}
\addcontentsline{toc}{section}{References}
\begin{thebibliography}{99}

\bibitem{LRR}
Barcel\'o C, Liberati S and Visser M 2005 {\it Living Rev. Rel.} \href{http://relativity.livingreviews.org/Articles/lrr-2005-12/}{{\bf 8} 12}
\bibitem{unruh81} Unruh W G 1981 {\it Phys. Rev. Lett.} \href{http://prl.aps.org/abstract/PRL/v46/i21/p1351_1}{{\bf 46} 1351}
\bibitem{Tachnion} Lahav O, Itah A, Blumkin A, Gordon C and 
Steinhauer J 2009 A sonic black hole in a density-inverted Bose--Einstein condensate arXiv:\href{http://arxiv.org/abs/0906.1337}{0906.1337} [cond-mat.quant-gas]
\bibitem{hawking} Hawking S W 1974 {\it Nature} \href{http://www.nature.com/nature/journal/v248/n5443/abs/248030a0.html}{{\bf 248} 30} 
\bibitem{BEC}
Garay L J, Anglin J R, Cirac J I and Zoller P 2000 {\it Phys. Rev. Lett.} \href{http://prl.aps.org/abstract/PRL/v85/i22/p4643_1}{{\bf 85} 4643}
\bibitem{fermi} Giovanazzi S 2005 {\it Phys. Rev. Lett.} \href{http://prl.aps.org/abstract/PRL/v94/i6/e061302}{{\bf 94} 061302}
\bibitem{He} Jacobson T A and Volovik G E 1998 {\it Phys. Rev.} D \href{http://prd.aps.org/abstract/PRD/v58/i6/e064021}{{\bf 58} 064021} 
\bibitem{slow}
Leonhardt U and Piwnicki P 2000 {\it Phys. Rev. Lett.} \href{http://prl.aps.org/abstract/PRL/v84/i5/p822_1}{{\bf 84} 822} 

\hspace{-13pt}
Reznik B 2000 {\it Phys. Rev. D} \href{http://prd.aps.org/abstract/PRD/v62/i4/e044044}{{\bf 62} 044044}

\hspace{-13pt}
Unruh W G and Sch\"utzhold R 2003 {\it Phys. Rev. D} \href{http://prd.aps.org/abstract/PRD/v68/i2/e024008}{{\bf 68} 024008}
\bibitem{waveguide}
Sch\"utzhold R and Unruh W G 2005 {\it Phys. Rev. Lett.} \href{http://prl.aps.org/abstract/PRL/v95/i3/e031301}{{\bf 95} 031301}

\hspace{-13pt}
Philbin T G, Kuklewicz C, Robertson S, Hill S, Konig F and Leonhardt U 2008 {\it Science} \href{http://www.sciencemag.org/cgi/content/abstract/319/5868/1367}{{\bf 319} 1367}
\bibitem{ions} Horstmann B, Reznik B, Fagnocchi S and Cirac J I 2009 Hawking Radiation from an Acoustic Black Hole on an Ion Ring arXiv:\href{http://arxiv.org/abs/0904.4801}{0904.4801} [quant-ph]
\bibitem{P-G} Visser M 1998 {\it Class. Quant. Grav.} \href{http://iopscience.iop.org/0264-9381/15/6/024/}{{\bf 15} 1767} 
\bibitem{superrad} Slatyer T R and Savage C M 2005 {\it Class. Quantum. Grav.} \href{http://iopscience.iop.org/0264-9381/22/19/002/}{{\bf 22} 3833} 
\bibitem{zeldovich}
Zel'dovich Y B 1972 in \emph{Magic without magic} ed J R Klauder (San Francisco: Freeman) pp~277--288
\bibitem{silke}
Visser M and Weinfurtner S E C 2005 {\it Class. Quant. Grav.} \href{http://iopscience.iop.org/0264-9381/22/12/011/}{{\bf 22} 2493} 
\bibitem{expandinguniverse}
Barcel\'o C, Liberati S and Visser M 2003 {\it Int. J. Mod. Phys.}  D \href{http://www.worldscinet.com/ijmpd/12/1209/S0218271803004092.html}{{\bf 12} 1641}

\hspace{-13pt}
Fedichev P O and Fischer U R 2003 {\it Phys. Rev. Lett.} \href{http://prl.aps.org/abstract/PRL/v91/i24/e240407}{{\bf 91} 240407}

\hspace{-13pt}
Weinfurtner S, Jain P, Visser M and Gardiner C W 2009 {\it Class. Quant. Grav.} \href{http://iopscience.iop.org/0264-9381/26/6/065012/}{{\bf 26} 065012}

\hspace{-13pt}
Jain P, Weinfurtner S, Visser M and Gardiner C W 2007
Analogue model of a FRW universe in Bose-Einstein condensates: Application of the classical field method arXiv:\href{http://arxiv.org/abs/0705.2077}{0705.2077} [cond-mat.other]
\bibitem{HR-robust}
Unruh W G and Schutzhold R 2005 {\it Phys. Rev.} D \href{http://prd.aps.org/abstract/PRD/v71/i2/e024028}{{\bf 71} 024028}

\hspace{-13pt}
Barcel\'o C, Garay L J and Jannes G 2009 {\it Phys. Rev.} D \href{http://prd.aps.org/abstract/PRD/v79/i2/e024016}{{\bf 79} 024016}

\hspace{-13pt}
Macher J and Parentani R 2009 {\it Phys. Rev.} D \href{http://prd.aps.org/abstract/PRD/v79/i12/e124008}{{\bf 79} 124008}

\hspace{-13pt}
Balbinot R, Fabbri A, Fagnocchi S and Parentani R 2005 {\it Riv. Nuovo Cim.} {\bf 28} 1
\bibitem{corr-BEC}  
Balbinot R, Fabbri A, Fagnocchi S, Recati A and Carusotto I 2008  {\it Phys. Rev.} A \href{http://pra.aps.org/abstract/PRA/v78/i2/e021603}{{\bf 78} 021603}

\hspace{-13pt}
Carusotto I, Fagnocchi S, Recati A, Balbinot R and Fabbri A 2008 {\it New J. Phys.} \href{http://iopscience.iop.org/1367-2630/10/10/103001/}{{\bf 10} 103001}

\hspace{-13pt}
Recati A, Pavloff N and Carusotto I 2009 Bogoliubov Theory of acoustic Hawking radiation in Bose--Einstein Condensates arXiv:\href{http://arxiv.org/abs/0907.4305}{0907.4305} [cond-mat.quant-gas]
\bibitem{macherparentani}
Macher J and Parentani R 2009 {\it Phys. Rev.} A \href{http://pra.aps.org/abstract/PRA/v80/i4/e043601}{{\bf 80} 043601}
\bibitem{toymodel}
Liberati S, Visser M and Weinfurtner S 2006 {\it Class. Quant. Grav.} \href{http://iopscience.iop.org/0264-9381/23/9/023/}{{\bf 23} 3129}

\hspace{-13pt}
Girelli F, Liberati S and Sindoni L 2008 {\it Phys. Rev.} D \href{http://prd.aps.org/abstract/PRD/v78/i8/e084013}{{\bf 78} 084013}

\hspace{-13pt}
Girelli F, Liberati S and Sindoni L 2009 {\it Phys. Rev.} D \href{http://prd.aps.org/abstract/PRD/v79/i4/e044019}{{\bf 79} 044019}

\hspace{-13pt}
Skakala J and Visser M 2009 {\it J. Phys. Conf. Ser.} \href{http://iopscience.iop.org/1742-6596/189/1/012037/}{{\bf 189} 012037}
\bibitem{BLH}
Hu B L 2005 {\it Int. J. Theor. Phys.} \href{http://www.springerlink.com/content/7150m252p5394360/}{{\bf 44} 1785}
\bibitem{Bombelli:1987aa}
Bombelli L, Lee J H, Meyer D and Sorkin R 1987 {\it Phys. Rev. Lett.} \href{http://prl.aps.org/abstract/PRL/v59/i5/p521_1}{{\bf 59} 521}

\hspace{-13pt}
Brightwell G, Henson J and Surya S 2009 {\it J. Phys. Conf. Ser.} \href{http://iopscience.iop.org/1742-6596/174/1/012049/}{{\bf 174} 012049}
\bibitem{Oriti:2007qd}
Oriti D 2007 Group field theory as the microscopic description of the quantum spacetime fluid: a new perspective on the continuum in quantum gravity arXiv:\href{http://arxiv.org/abs/0710.3276}{0710.3276} [gr-qc]

\hspace{-13pt}
Girelli F, Livine E R and Oriti D 2009 4d Deformed Special Relativity from Group Field Theories arXiv:\href{http://arxiv.org/abs/0903.3475}{0903.3475} [gr-qc]
\bibitem{Konopka:2008hp}
Konopka T 2008 Statistical Mechanics of Graphity Models arXiv:\href{http://arxiv.org/abs/0805.2283}{0805.2283} [hep-th]

\hspace{-13pt}
Konopka T, Markopoulou F and Severini S 2008 {\it Phys. Rev.} D \href{http://prd.aps.org/abstract/PRD/v77/i10/e104029}{{\bf 77} 104029}

\hspace{-13pt}
Konopka T, Markopoulou F and Smolin L 2006 Quantum graphity arXiv:\href{http://arxiv.org/abs/hep-th/0611197}{hep-th/0611197}
\bibitem{Dreyer}
Dreyer O 2007 Why things fall {\it Proc. From Quantum to Emergent Gravity: Theory and Phenomenology, Trieste, Italy, 11-15 Jun 2007}

\hspace{-13pt}
Dreyer O 2006 Emergent general relativity  {\it Towards Quantum Gravity} ed D Oriti (Cambridge: Cambridge University Press) [arXiv:\href{http://arxiv.org/abs/gr-qc/0604075}{gr-qc/0604075}]
\bibitem{Padma}
Padmanabhan T 2009
A Dialogue on the Nature of Gravity arXiv:\href{http://arxiv.org/abs/0910.0839}{0910.0839} [gr-qc]
\bibitem{SIGRAV}
Liberati S, Girelli F and Sindoni L 2009
Analogue Models for Emergent Gravity arXiv:\href{http://arxiv.org/abs/0909.3834}{0909.3834} [gr-qc]
\bibitem{chinese}
Ge X H and Sin S J 2010 arXiv:\href{http://arxiv.org/abs/1001.0371}{1001.0371} [hep-th]
\bibitem{haber81}
Haber H E and Weldon H A 1981 {\it Phys. Rev. Lett.} \href{http://prl.aps.org/abstract/PRL/v46/i23/p1497_1}{{\bf 46} 1497}

\hspace{-13pt}
Haber H E and Weldon H A 1982 {\it Phys. Rev. D} \href{http://prd.aps.org/abstract/PRD/v25/i2/p502_1}{{\bf 25} 502}
\bibitem{kapusta81}
Kapusta J I 1981 {\it Phys. Rev.} D \href{http://prd.aps.org/abstract/PRD/v24/i2/p426_1}{{\bf 24} 426}
\bibitem{singh84}
Singh S and Pathria R K 1984 {\it Phys. Rev.} A \href{http://pra.aps.org/abstract/PRA/v30/i1/p442_1}{{\bf 30} 442}

\hspace{-13pt}
Singh S and Pathria R K 1984 {\it Phys. Rev.} A \href{http://pra.aps.org/abstract/PRA/v30/i6/p3198_1}{{\bf 30} 3198}
\bibitem{bernstein91}
Bernstein J and Dodelson S 1991 {\it Phys. Rev. Lett.} \href{http://prl.aps.org/abstract/PRL/v66/i6/p683_1}{{\bf 66} 683}
\bibitem{grether07}
Grether M, de Llano M and Baker G A 2007 {\it Phys. Rev. Lett.} \href{http://prl.aps.org/abstract/PRL/v99/i20/e200406}{{\bf 99} 200406}
\bibitem{witkowska09}
Witkowska E, Zin P and Gajda M 2009 {\it Phys. Rev.} D \href{http://prd.aps.org/abstract/PRD/v79/i2/e025003}{{\bf 79} 025003}
\bibitem{stringari03}
Pitaevskii L P and Stringari S 2003 {\it Bose--Einstein Condensation} (Oxford: Clarendon Press)
\bibitem{pethick08}
Pethick C J and Smith H 2008 {\it Bose--Einstein Condensation in Dilute Gases} (Cambridge: Cambridge University Press)
\bibitem{andersen07} 
Andersen J O 2007 {\it Phys. Rev.} D \href{http://dx.doi.org/10.1103/PhysRevD.75.065011}{{\bf 75} 065011}
\bibitem{Molina-Visser}
Visser M and Molina--Par{\'\i}s C 2010 Acoustic geometry for general relativistic barotropic irrotational fluid arXiv:\href{http://arxiv.org/abs/1001.1310}{1001.1310} [gr-qc]
\bibitem{Mattingly}
Mattingly D 2005 {\it Living Rev. Rel.} \href{http://relativity.livingreviews.org/Articles/lrr-2005-5/index.html}{{\bf 8} 5}
\bibitem{GAC}
Amelino-Camelia G 2008 Quantum Gravity Phenomenology arXiv:\href{http://arxiv.org/abs/0806.0339}{0806.0339} [gr-qc]
\bibitem{MacLib}
Liberati S and Maccione L 2009 {\it Ann. Rev. Nucl. Part. Sci.} \href{http://www.annualreviews.org/doi/abs/10.1146/annurev.nucl.010909.083640}{{\bf 59} 245}
\bibitem{Collins}
Collins J, Perez A, Sudarsky D, Urrutia L and Vucetich H 2004 {\it Phys. Rev. Lett.} \href{http://prl.aps.org/abstract/PRL/v93/i19/e191301}{{\bf 93} 191301}
\bibitem{IRS}
Iengo R, Russo J G and Serone M 2009 {\it J. High Energy Phys.} \href{http://iopscience.iop.org/1126-6708/2009/11/020/}{{\bf 0911} 020}
\end{thebibliography}
\end{document}